\documentclass{article}  

\usepackage{amsmath}
\usepackage{amsfonts}
\usepackage{amssymb}
\usepackage{amsthm}
\usepackage{mathtools}
\usepackage{mathrsfs}
\usepackage{mathpazo}

\usepackage{graphicx}
\usepackage{enumerate,color}
\usepackage{microtype}
\usepackage[margin=15pt,font=small,labelfont={bf,sf}]{caption}
\usepackage{sectsty}
\allsectionsfont{\sffamily}

\usepackage{tikz}
\usepackage[american]{circuitikz}
\usepackage{pgfplots}
\usetikzlibrary{calc}
\usetikzlibrary{shapes}
\usetikzlibrary{arrows}
\usetikzlibrary{decorations.markings}
\usetikzlibrary{decorations.pathmorphing}
\usetikzlibrary{positioning}
\tikzset{every picture/.style=thick}

\usepackage{hyperref}
\usepackage[nameinlink]{cleveref}

\definecolor{lightblue}{rgb}{0.54, 0.81, 0.94}
\hypersetup{
    colorlinks = true,
    linkcolor = lightblue,
    citecolor = lightblue,
    urlcolor = lightblue,
    }
    
\crefformat{equation}{(#2#1#3)}
\crefmultiformat{equation}{(#2#1#3)}%
{ and~(#2#1#3)}{, (#2#1#3)}{ and~(#2#1#3)}

\usepackage{amsmath}
\usepackage{amsfonts}
\usepackage{mathrsfs}
\usepackage{mathtools}

\newcommand{\abs}[1]{\ensuremath{\left\vert#1\right\vert}}

\newcommand{\C}{\ensuremath{\mathbb{C}}}
\DeclarePairedDelimiter{\ceil}{\lceil}{\rceil}

\newcommand{\diag}[1]{\mathrm{diag}\,\ensuremath{\left(#1\right)}}
\newcommand{\funof}[1]{\ensuremath{\left(#1\right)}}

\newcommand{\jw}{\ensuremath{\left(j\omega\right)}}
\newcommand{\jwo}{\ensuremath{\left(j\omega_0\right)}}

\newcommand{\Nat}{\ensuremath{\mathbb{N}}}
\newcommand{\norm}[1]{\ensuremath{\left\Vert #1 \right\Vert}}
\newcommand{\R}{\ensuremath{\mathbb{R}}}
\newcommand{\Rat}{\ensuremath{\mathscr{R}}}
\newcommand{\RH}{\ensuremath{\mathscr{RH}_\infty}}

\newcommand{\s}{\ensuremath{\left(s\right)}}

\newcommand\scalemath[2]{\scalebox{#1}{\mbox{\ensuremath{\displaystyle #2}}}}

\newtheorem{theorem}{Theorem}

\newtheorem{remark}{Remark}
\newtheorem{lemma}{Lemma}

\begin{document}

\title{Exploiting Heterogeneity in the Decentralised Control of Platoons}
\author{Richard Pates\thanks{The author is a member of the ELLIIT Strategic Research Area at Lund University. This work was supported by the ELLIIT Strategic Research Area. This project has received funding from ERC grant agreement No 834142.}}

\maketitle

\begin{abstract}
This paper investigates the use of decentralised control architectures with heterogeneous dynamics for improving performance in large-scale systems. Our focus is on two well-known decentralised approaches; the ‘predecessor following’ and ‘bidirectional’ architectures for vehicle platooning. The former, utilising homogeneous control dynamics, is known to face exponential growth in disturbance amplification throughout the platoon, resulting in poor scalability properties. We demonstrate that by incorporating heterogeneous control system dynamics, this limitation disappears entirely, even under bandwidth constraints. Furthermore, we reveal that introducing heterogeneity in the bidirectional architecture allows the platoon’s behaviour to be rendered independent of its length, allowing for highly scalable performance.
\end{abstract}


\section*{Notation}

Let $\Rat$ denote the set of proper real-rational functions in the indeterminate $s$. A given $p\in\Rat$ is said to be stable if it has bounded H-infinity norm
\[
\norm{p}_\infty{}=\sup_{s:\text{Re}\funof{s}>0}\abs{p\s},
\]
and the set of all stable real-rational functions will be denoted by $\RH$. Given $p\in\Rat$ and $c\in\Rat$, we say that $c$ internally stabilises $p$ if the four transfer functions
\[
\frac{1}{1+pc},\frac{p}{1+pc},\frac{c}{1+pc},\;\text{and}\;\frac{pc}{1+pc},
\]
are all in $\RH{}$.

\section{Introduction}

Subsystem dynamics and control system architectures are two crucial ingredients in describing the behaviour of large-scale systems. For example they have been shown to play a fundamental role in the emergence of coherence in platooning and flocking problems \cite{BJM12}. Similar behaviours have been observed in a variety of man-made networks, such as interarea oscillations in power systems \cite{Kun94}. Deepening our understanding of these phenomena holds the potential to inform improved methods for control system design in the large-scale setting.

Significant progress has been made within the control community in our understanding of networks constructed from components with identical dynamics. Here robust stability analyses \cite{OFM07,HHS09} and their interplay with network topology \cite{TBS23} have been key themes. However, the case of heterogeneous components is less well understood. While strong tools now exist for addressing, for instance, robust stability questions \cite{LV06,OSC08}, the prevailing perspective has been to treat heterogeneity as a complication to be mitigated or circumvented. This largely overlooks the potential benefits and opportunities that heterogeneous dynamics can offer, especially in the domain of scalability and disturbance attenuation. In contrast, this paper aims to highlight the advantages of heterogeneous dynamics, particularly in the design of decentralised controllers.

The first theme we explore is that of mistuning an ensemble of controllers to disrupt disturbance propagation. This is similar in spirit to previous works on vehicle platoons, where breaking symmetry in bidirectional architectures was shown to enhance performance \cite{BMH09}. We adopt a more classical control perspective centred on misaligning peaks in frequency responses, illustrating its capability to counteract the poor scalability properties of the ‘predecessor following’ control architecture for platoons \cite{SPH04}.

The second theme revolves around tweaking the dynamics of a few control laws to significantly alter system-wide performance. This concept has parallels with \cite{CPS22}, where a few carefully tuned agents were employed to disrupt traffic bottlenecks and alleviate traffic congestion. The tools we develop are perhaps more similar to the use of impedance matching for transmission lines and tunable mass dampers for buildings. We demonstrate their efficacy in rendering the bidirectional architecture for platoons essentially scale-invariant.

The remainder of the paper is structured as follows: In \cref{sec:prob}, we introduce the dynamical models for the ‘predecessor following’ and ‘bidirectional’ architectures. In \cref{sec:resa}, we discuss heterogeneous design in the predecessor following case. Our main result, presented as \Cref{thm:1}, demonstrates that the exponential growth in disturbance amplification inherent in the design of homogeneous control laws vanishes entirely when decentralised controllers are allowed. Finally, in \cref{sec:resb}, we examine the bidirectional case, developing heterogeneous designs through a curious factorisation result, presented as \Cref{lem:1}, enabling platoon behaviour that is essentially independent of its length and ensuring highly scalable performance guarantees.

\section{Problem formulation}\label{sec:prob}

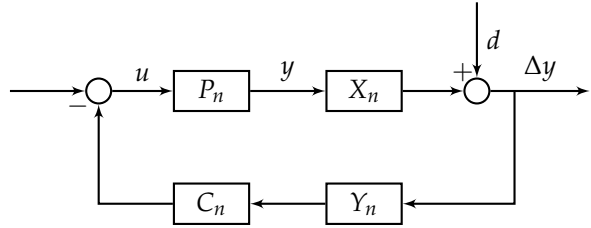
\begin{figure}[h]
\vspace{.2cm}
\centering
\tikzstyle{block} = [draw, rectangle, minimum height=0.5cm, minimum width=1cm]
\tikzstyle{sum} = [draw, circle, node distance=1.5cm]
\tikzstyle{input} = [coordinate]
\tikzstyle{output} = [coordinate]
\tikzstyle{disturbance} = [coordinate]
\tikzstyle{pinstyle} = [pin edge={to-,thin,black}]
\begin{tikzpicture}[auto, node distance=1.5cm,>=latex']
    \node [input, name=input] {};
    \node [sum, right = 1cm of input] (sum1) {};
    \node [block, right of=sum1] (BT) {$P_n$};
    \node [block, right =1cm of BT] (plant) {$X_n$};
    \node [sum, right of=plant] (sum2) {};
    \node [output, right of=sum2] (output){};
    \node [disturbance, above = 1cm of sum2] (disturbance){};
    \node [block, below of=plant] (B){$Y_n$};
    \node [block, below of=BT] (controller){$C_n$};

    \draw [draw,->] (input) --  (sum1);
    \draw [->] (sum1) --node[above] {$u$} (BT);
    \draw [->] (BT) --node[above] {$y$} (plant);
    \draw [->] (plant) -- node[pos=0.99]{$+$}(sum2);
    \draw [->] (sum2) -- node[name=y]{$\Delta{}y$}(output);
    \draw [->] (disturbance)  --node{$d$}(sum2);
    \draw [->] (sum2)+(0.5,0) |- (B);
    \draw [->] (B) -- (controller);
    \draw [->] (controller) -| node[pos=0.99]{$-$}(sum1);
 \end{tikzpicture}
 \caption{The feedback configuration} 
 \label{fig:block}
\end{figure}

We consider the linear time-invariant system depicted in \Cref{fig:block}. The blocks labelled $P_n$ and $C_n$ denote diagonal matrices of rational functions
\[
P_n=\diag{p_1,\ldots{},p_n}\;\text{and}\;C_n=\diag{c_1,\ldots{},c_n},
\]
and $X_n\in\R^{n\times{}n}$ and $Y_n\in\R^{n\times{}n}$. The idea behind this interconnection is that the rational functions $p_1,\ldots{},p_n$ describe the dynamics of a set of $n$ processes that we wish to control, and $c_1,\ldots{},c_n$ a set of decentralised control laws to be designed. The matrices $X_n$ and $Y_n$ describe the combination of outputs that are available for each controller to act on, and how the control signals are then distributed out again to the processes, respectively. 

We are interested in investigating the role of heterogeneity in controller dynamics, and so restrict our attention to arguably the simplest interesting choices for $P_n,X_n$ and $Y_n$. More specifically we suppose that $p_k=s^{-m}$ for $k=1,\ldots{},n$,
\[
X_n
=\scalemath{.8}{ \begin{bmatrix}
        1 & 0 & \cdots{} &\cdots&0 \\
        -1 & \ddots{} & \ddots{} &\ddots{} &\vdots{}\\
        0 & \ddots & \ddots & \ddots&\vdots{}  \\
        \vdots{} & \ddots{} & \ddots{}& \ddots & 0\\
        0 & \cdots{} & 0 & -1 &1
        \end{bmatrix} }\in \R^{n\times n},
\]
and either $Y_n=I$ or $Y_n=X_n^{\top{}}$. The motivation for this setting comes from the vehicle platooning literature, where the ‘vehicles' have $m$th order integrator dynamics (a simple Newton's law model would correspond to $m=2$). The given $X_n$ matrix then specifies that each controller can act on the difference between the positions of two neighbouring vehicles (i.e. for $k>1$, $\Delta{}y_k=y_k-y_{k-1}+d_k$). Note that the slight asymmetry in this description leading to $\Delta{}y_1=y_1+d_1$ is also fairly standard, and the external signal $d_1$ usually include the position of a lead vehicle so that $c_1$ is in effect acting on the difference in position between the first car and this leader. The two choices of $Y_n$ then correspond to two well studied decentralised control architectures.
\begin{enumerate}
\item If $Y_n=I$, \textit{predecessor following} is employed, in which case each vehicle's control input is selected to regulate the distance between itself and the vehicle in front.
\item If $Y_n=X_n^{\top{}}$, a \textit{bidirectional} architecture is employed, in which case each vehicle's control input is selected to regulate the mismatch between its distance from the vehicle in front and the vehicle behind.
\end{enumerate}
For further discussions of these architectures, and a good entry point to the platooning literature, see \cite{SPH04}.

\section{Results}

\subsection{Predecessor following}\label{sec:resa}

\begin{figure*}[h]

\begin{center}
\tikzstyle{block} = [draw, rectangle, minimum height=0.5cm, minimum width=1cm]
\tikzstyle{sum} = [draw, circle, node distance=1cm]
\tikzstyle{input} = [coordinate]
\tikzstyle{output} = [coordinate]
\tikzstyle{disturbance} = [coordinate]
\tikzstyle{pinstyle} = [pin edge={to-,thin,black}]
\begin{tikzpicture}[auto, node distance=1.5cm,>=latex']
    \node [input] (I1) {};
    \node [sum, right = .6cm of I1] (S1) {};
    \node [block, right =.3cm of S1] (C1) {$c_1$};
    \node [block, right =.6cm of C1] (P1) {$p_1$};
    \node [output, right = .3cm of P1] (O1){};
    \node [output, below = .8cm of P1] (O2){};

     \draw [->] (I1) --node[above] {$-d_1$} (S1);
     \draw [->] (S1) -- (C1);
     \draw [->] (C1) --node[above] {$u_1$} (P1);
     \draw[-]  (O1) |- (O2);
     \draw[->]  (O2) -| node[pos=0.99]{$-$}(S1);

    \node [sum, right = 1cm of P1] (S2) {};
    \node [input, above = .4cm of S2] (I2) {};
    \node [block, right =.6cm of S2] (C2) {$c_2$};
    \node [block, right =.6cm of C2] (P2) {$p_2$};
     \node [output, right = .3cm of P2] (O3){};
    \node [output, below = .8cm of P2] (O4){};
    \node [output, right = 1cm of P2] (O5){};
    
    \draw [->] (P1) --node[above] {$y_1$} (S2);
    \draw[-]  (O3) |- (O4);
    \draw[->]  (O4) -| node[pos=0.99]{$-$}(S2);
    
     \draw [->] (I2) --node[above,yshift=.1cm] {$-d_2$} (S2);
     \draw [->] (S2) -- (C2);
     \draw [->] (C2) --node[above] {$u_1$} (P2);
     \draw [->] (P2) --node[above] {$y_2$} (O5);

    \node [output, right = .2cm of O5] (OO1){};
    \node [output, right = .4cm of O5] (OO2){};
    \node [output, right = .6cm of O5] (OO3){};
    
    \draw[black,fill=black] (OO1) circle (.3ex);
    \draw[black,fill=black] (OO2) circle (.3ex);
    \draw[black,fill=black] (OO3) circle (.3ex);
    
 \end{tikzpicture}
 \end{center}
 \caption{The predecessor  follower architecture.} 
 \label{fig:block1}
\end{figure*}
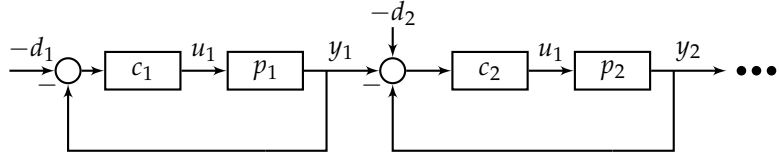

We begin with an analysis of the predecessor following architecture. This setting is particularly simple, and allows the block diagram in \Cref{fig:block} to be unwound into the equivalent cascaded interconnection illustrated in \Cref{fig:block1}. This reveals that each controller $c_k$ appears in its own decoupled feedback loop, and that the complementary sensitivity functions
\[
\frac{c_kp_k}{1+c_kp_k}
\]
play a key in role in describing how disturbances spread through the platoon. For example, the transfer function from a disturbance applied to the first vehicle $d_1$ to the position of the final vehicle $y_n$ is given by
\begin{equation}\label{eq:amp}
y_n=-\funof{\prod_{k=1}^n\frac{c_kp_k}{1+c_kp_k}}d_1.
\end{equation}
The scalability properties of this architecture when the controllers are identical are much maligned, and a particularly elegant refutation is given in \cite[Theorem 1]{SPH04}. We repeat the key steps because they are relevant here. 

First recall that $p_k=s^{-m}$, and note that if identical control laws $c=c_1=c_2,\ldots{}$ are used, then for all $k$,
\[
\frac{c_kp_k}{1+c_kp_k}=\frac{c\s}{s^m+c\s}.
\]
Middleton's complementary sensitivity integral \cite{Mid91} then ensures that whenever $m\geq{}2$,
\[
\int_0^\infty{}\ln\abs{\frac{c\jw}{\funof{j\omega}^m+c\jw}}\frac{1}{\omega^2}\,\mathrm{d}\omega\geq{}0.
\]
Since $\lim_{s\rightarrow{}\infty}\tfrac{c\s}{s^m+c\s}=0$, it then follows that there must exist a frequency $\omega_0>0$ such that
\[
\abs{\frac{c\jwo}{\funof{j\omega_0}^m+c\jwo}}>1.
\]
Comparison with \cref{eq:amp} then reveals that 
\[
\abs{\frac{y_n\jwo}{d_1\jwo}}\geq{}\abs{\frac{c\jwo}{\funof{j\omega_0}^m+c\jwo}}^n,
\]
and so disturbances become exponentially amplified as the length of the platoon is increased. This lack of scalability whenever the process has double integrator dynamics or higher may be bluntly stated as follows (to see the result for $m=1$, set $c=1$):
\begin{equation}\label{eq:siso}
\inf_{c\in\mathscr{S}}\sup_{n\in\Nat}\norm{\funof{\frac{c}{s^m+c}}^n}_\infty
=
\begin{cases}
1&\text{if $m=1$,}\\
\infty{}&\text{otherwise.}
\end{cases}
\end{equation}
It is natural then to ask, \textit{‘‘to what extent can heterogeneity help the situation?''}. Clearly it can, since the rate of growth of \cref{eq:amp} was compounded by the fact that every complementary sensitivity function had its peak at the same frequency. However perhaps even a heterogeneous design will suffer from the same lack of scalability eventually. One may consider this especially likely when considering a maximum bandwidth assumption on the control (which would always hold in practice, and are known to impose more severe fundamental limitations, c.f. \cite[p.91]{DFT91}) along the lines of
\begin{equation}\label{eq:bandwidth}
\abs{\funof{\frac{p_kc_k}{1+p_kc_k}}\jw}\leq{}1\,\text{for all $\omega\geq{}\omega_b>0$.}
\end{equation}
The following theorem demonstrates that this is not the case, by showing that in the heterogeneous analogue of \cref{eq:siso}, an infimal value of 1 can be achieved for all $m\in\Nat$, independently of any bandwidth like constraint on the form of \cref{eq:bandwidth}.

\begin{theorem}\label{thm:1}
Let $m\in\Nat$, $\omega_{\mathrm{bw}}>0$, and $\mathscr{S}_{\mathrm{bw}}$ be the set of control laws $c\in\Rat$ that internally stabilise $s^{-m}$ and achieve
\[
\abs{\frac{c\jw}{\funof{j\omega}^m+c\jw}}\leq{}1\;\text{for all $\omega\geq{}\omega_{\text{bw}}$.}
\]
Then
\begin{equation}\label{eq:mimo}
\inf_{c_1,c_2,\ldots{}\in\mathscr{S}_{\mathrm{bw}}}\sup_{n\in\Nat}\norm{\prod_{k=1}^n\frac{c_k}{s^m+c_k}}_\infty
=
1.
\end{equation}
\end{theorem}


\begin{proof}
Throughout the proof we will need to evaluate various algebraic combinations of rational functions at particular values of $s\in\C$. To avoid defining too many intermediate functions, we will often denote such evaluations using notation along the lines of
\[
\funof{\frac{1}{1+pc}}\funof{0}.
\]
This should be read as evaluating the rational function $\tfrac{1}{1+pc}$ at the point $s=0$. 

We now proceed with the proof. Let $p=s^{-m}$ and $\mathscr{S}$ denote the set of control laws that internally stabilise $p$ (we introduce this redundant notation both because $p$ is easier to type than $s^{-m}$, and also to discuss generalisations in \Cref{rem:1}, which follows the proof). We start by considering the simpler optimal control problem 
\begin{equation}\label{eq:singopt}
\inf_{c\in\mathscr{S}}\norm{\frac{pc}{1+pc}}_\infty.
\end{equation}
Since for any $c\in\mathscr{S}$
\[
\funof{\frac{pc}{1+pc}}\funof{0}=1,
\]
it is clear that the infimal value of \cref{eq:singopt} is at least 1. We will now show that for any $\varepsilon>0$ it is possible to find a $p\in\mathscr{S}$ with the following two properties:
\begin{enumerate}
\item $\norm{\tfrac{pc}{1+pc}}_\infty{}\leq{}1+\varepsilon{}$;
\item $\abs{\funof{\tfrac{pc}{1+pc}}\jw}\leq{}1$ for all $0\leq{}\omega{}\leq{}\underline{\omega}$, where $\underline{\omega}>0$.
\end{enumerate}
The first property shows that the infimal value of \cref{eq:singopt} is 1.The second property will allow us to extend the found solution for \cref{eq:singopt} to show that \cref{eq:mimo} achieves the same infimal value. We will discover along the way that the bandwidth constraint imposes no restriction.

\textit{Part I: Constructing control laws meeting 1) and 2)}: We first show that it is sufficient to consider the case that $m$ is divisible by 4. To see this, assume that there exist control laws $\bar{c}_{\varepsilon,\ell}$ that meet 1) and 2) whenever $\ell=4,8,12,\ldots{}$, and define
\[
c_{\varepsilon,m}=s^{-\ell+m}\bar{c}_{\varepsilon,\ell}
\]
for all $m\in\Nat$, where $\ell=4\ceil{m/4}$. It follows from the above that
\begin{equation}\label{eq:stab1}
\frac{p}{1+pc_{\varepsilon,m}}=s^{\ell-m}\frac{s^{-\ell}}{1+s^{-\ell}\bar{c}_{\varepsilon,\ell}}
\end{equation}
and
\begin{equation}\label{eq:stab2}
\frac{c_{\varepsilon,m}}{1+pc_{\varepsilon,m}}=s^{m}\frac{s^{-\ell}\bar{c}_{\varepsilon,\ell}}{1+s^{-\ell}\bar{c}_{\varepsilon,\ell}}.
\end{equation}
Since by assumption $\bar{c}_{\varepsilon,\ell}$ internally stabilises $s^{-\ell}$,
\[
\frac{s^{-\ell}}{1+s^{-\ell}\bar{c}_{\varepsilon,\ell}},\frac{s^{-\ell}\bar{c}_{\varepsilon,\ell}}{1+s^{-\ell}\bar{c}_{\varepsilon,\ell}}\in\RH,
\]
and therefore by \cref{eq:stab1,eq:stab2}, $c_{\varepsilon,m}$ internally stabilises $p$. Furthermore since
\[
\frac{pc_{\varepsilon,m}}{1+pc_{\varepsilon,m}}=\frac{s^{-\ell}\bar{p}_{\varepsilon,\ell}}{1+s^{-\ell}\bar{c}_{\varepsilon,\ell}},
\]
$c_{\varepsilon,m}$ also meets 1) and 2).

We now assume that $m$ is divisible by 4, and show that a $c$ meeting 1) and 2) exists. We will take a model matching approach (we follow the treatment in \cite[\S{}9]{DFT91}, and so introduce $M,N,X,Y\in\RH$ such that
\begin{equation}\label{eq:cop}
NX+MY=1,\,p=N/M.
\end{equation}
The Youla parametrisation of all stabilising control laws then gives that $c\in\mathscr{S}$ if and only if
\[
c=\frac{X+MQ}{Y-NQ}
\]
for some $Q\in\RH$. Using the above to eliminate $C$ shows that \cref{eq:singopt} is equivalent to
\[
\inf_{Q\in\RH}\norm{N\funof{X+MQ}}_\infty{}.
\]
Now let $Q=-XQ_1$, where $Q_1$ is some as yet unspecified function in $\RH$. Fix $N$ and $M$ according to
\[
N = \frac{1}{\funof{s+1}^m},M = \frac{s^m}{\funof{s+1}^m}.
\]
Since
\[
\begin{aligned}
\funof{1+s}^{2m}&=\sum_{k=0}^{2m}\binom{2m}{k}s^k\\
&=\funof{\sum_{k=0}^{m-1}\binom{2m}{k}s^k}+s^m\funof{\sum_{l=m}^{2m}\binom{2m}{l}s^{l-m}},
\end{aligned}
\]
we see from \cref{eq:cop} that
\[
X = \frac{\sum_{k=0}^{m-1}\binom{2m}{k}s^m}{\funof{s+1}^m},Y=\frac{\sum_{l=m}^{2m}\binom{2m}{l}s^{l-m}}{\funof{s+1}^m}.
\]
Therefore
\[
N\funof{X+MQ}=\tfrac{\sum_{k=0}^{m-1}\binom{2m}{k}s^m}{\funof{s+1}^{2m}}\funof{1-\tfrac{s^m}{\funof{s+1}^m}Q_1}.
\]
Forming the Maclaurin series of $NX$ yields
\begin{equation}\label{eq:mac1}
NX=\frac{\sum_{k=0}^{m-1}\binom{2m}{k}s^m}{\funof{s+1}^{2m}}=1-\binom{2m}{m}s^m+\ldots{}
\end{equation}
Since $m$ is divisible by 4, $\funof{j\omega}^m=\omega^m$. When combined with the above, this implies that for sufficiently small $\omega$, $\abs{\funof{NX}\jw}<1$. Therefore there exists an ${\omega_1}>0$ such that for all $0\leq{}\omega\leq{}{\omega}_1$,
\begin{equation}\label{eq:ineq1}
\abs{\funof{NX}\jw}\leq{}1.
\end{equation}
We will now construct a $Q_1\in\RH$ such that
\begin{equation}\label{eq:ineq2}
\norm{\funof{1-\tfrac{s^m}{\funof{s+1}^m}Q_1}}_\infty\leq{}1+\varepsilon,
\end{equation}
and for all $\omega\geq{}{\omega}_1$,
\begin{equation}\label{eq:ineq3}
\abs{\funof{1-\tfrac{\omega^m}{\funof{j\omega+1}^m}Q_1\jw}}\leq{}\frac{1}{\norm{NX}_\infty}.
\end{equation}
This will be sufficient for 1) since
\[
\begin{aligned}
&\norm{N\funof{X+MQ}}_\infty{}=\sup_{\omega\in\R}\abs{\funof{NX\funof{1-MQ_1}}\jw},
\end{aligned}
\]
\cref{eq:ineq1,eq:ineq2} together ensure that for all $0\leq{}\omega\leq{}{\omega}_1$
\[
\abs{\funof{NX\funof{1-MQ_1}}\jw}\leq{}1+\varepsilon{},
\]
and \cref{eq:ineq3} ensures that for all $\omega\geq{}{\omega}_1$,
\[
\abs{\funof{NX\funof{1-MQ_1}}\jw}\leq{}1.
\]
Consider now
\[
Q_1 = \tfrac{\funof{s+1}^{m}}{\funof{s+\gamma_a}\funof{s+\gamma_b}^{m-1}},
\]
and so
\[
1-\tfrac{s^m}{\funof{s+1}^m}Q_1=1-\tfrac{s/\gamma_a}{s/\gamma_a+1}\tfrac{\funof{s/\gamma_b}^{m-1}}{\funof{s/\gamma_b+1}^{m-1}}.
\]
Observe now that since $\lim_{s\rightarrow{}0}\tfrac{s}{s+1}=0$, there exists an $\omega_a>0$ such that for all $0\leq{}\omega\leq{}\omega_a$,
\[
\abs{\tfrac{j\omega}{1+j\omega}}\leq{}\varepsilon.
\]
Therefore for all $0\leq{}\omega\leq{}\omega_a\gamma_a$,
\begin{align}
\nonumber{}\abs{1-\tfrac{\omega^m}{\funof{j\omega+1}^m}Q_1\jw}&\leq{}1+\abs{\tfrac{j\omega/\gamma_a}{j\omega/\gamma_a+q}}\abs{\tfrac{j\omega{}/\gamma_b}{\funof{j\omega{}/\gamma_b+1}}}^{m-1}\\
\label{eq:imp3}&\leq{}1+\varepsilon
\end{align}
Furthermore, since $\lim_{s\rightarrow{}\infty}\tfrac{s}{s+1}=1$, for any $\delta>0$ there exists an $\omega_b>0$ such that for all $\omega\geq{}\omega_b$,
\[
\abs{1-\tfrac{\funof{j\omega}^{m-1}}{\funof{j\omega+1}^{m-1}}}\leq{}\delta.
\]
Writing
\[
\begin{aligned}
1-&\tfrac{s/\gamma_a}{s/\gamma_a+1}\tfrac{\funof{s/\gamma_b}^{m-1}}{\funof{s/\gamma_b+1}^{m-1}}\\
&=\funof{1-\tfrac{s/\gamma_a}{s/\gamma_a+1}}+\tfrac{s/\gamma_a}{s/\gamma_a+1}\funof{1-\tfrac{\funof{s/\gamma_b}^{m-1}}{\funof{s/\gamma_b+1}^{m-1}}},\\
&=\tfrac{1}{s/\gamma_a+1}+\tfrac{s/\gamma_a}{s/\gamma_a+1}\funof{1-\tfrac{\funof{s/\gamma_b}^{m-1}}{\funof{s/\gamma_b+1}^{m-1}}},
\end{aligned}
\]
we then see that if we choose $\gamma_b>0$ such that $\gamma_b\leq{}\gamma_a\omega_a/\omega_b$, then for all $\omega\geq{}\gamma_a\omega_a$,
\begin{equation}\label{eq:imp4}
\abs{1-\tfrac{\omega^m}{\funof{j\omega+1}^m}Q_1\jw}\leq{}\tfrac{1}{\sqrt{\omega_a^2/\gamma_a^2+1}}+\delta.
\end{equation}
By choosing $\gamma_a>0$ and $\delta>0$ to be sufficiently small, it is possible to make both $\omega_a\gamma_a\leq{}\omega_1$ and
\[
\tfrac{1}{\sqrt{\omega_a^2/\gamma_a^2+1}}+\delta\leq{}\tfrac{1}{\norm{NX}_\infty{}}.
\]
It then follows that \cref{eq:imp3,eq:imp4} imply \cref{eq:ineq2,eq:ineq3}, and therefore there exists a $c$ such that 1) holds. Finally, forming the Maclaurin series
\[
1-\tfrac{s/\gamma_a}{s/\gamma_a+1}\tfrac{\funof{s/\gamma_b}^{m-1}}{\funof{s/\gamma_b+1}^{m-1}}=1-\tfrac{1}{\gamma_a\gamma_b^{m-1}}s^m
\]
and combining it with \cref{eq:mac1} shows that
\[
N\funof{X+MQ}=1-\funof{\binom{2m}{m}+\tfrac{1}{\gamma_a\gamma_b^{m-1}}}s^m+\ldots{}.
\]
Again observing that $\funof{j\omega}^m=\omega^m$ (and noting that the coefficient in front of $s^m$ is negative) shows that there must exist an $\underline{\omega}>0$ such that 2) holds.

\textit{Constructing the solution to \cref{eq:mimo}:} Our solution hinges on the following observations:
\begin{enumerate}
\item[a)]
\[
\begin{aligned}
\!\!\!\!\!\!\!\!\!\!\inf_{c_1,c_2,\ldots{}\in\mathscr{S}}\sup_{n\in\Nat}\norm{\prod_{k=1}^n\frac{pc_k}{1+pc_k}}_\infty&\geq{}\inf_{c_1\in\mathscr{S}}\norm{\frac{pc_1}{1+pc_1}}_\infty{},\\
&=1.
\end{aligned}
\]
\item[b)] Given any $p$, $c$, and $\gamma$,
\[
\funof{pc}\funof{\gamma{}s}=p\s{}\funof{\gamma^{-m}c\funof{\gamma{}s}}.
\]
\item[c)] Since $p$ is strictly proper, for any $c\in\mathscr{S}$,
\[
\lim_{s\rightarrow{}\infty{}}\funof{\frac{pc}{1+pc}}\s=0.
\]
\end{enumerate}
Let $\varepsilon>0$ and $c_\varepsilon\in\mathscr{S}$ meet 1) and 2) from the previous part. By c) there exists an $\overline{\omega}$ such that for all $\omega\geq{}\overline{\omega}$,
\[
\abs{\funof{\frac{pc_\varepsilon}{1+pc_\varepsilon}}\jw}\leq{}1.
\]
If we define 
\[
c_{\varepsilon{}}^\gamma=\gamma^{-m}c_{\varepsilon}\funof{\gamma{}s},
\]
it then follows from b) that
\begin{equation}\label{eq:sepfre}
\abs{\funof{\frac{pc_\varepsilon^\gamma}{1+pc_\varepsilon^\gamma}}\jw}\leq{}
\begin{cases}
1+\varepsilon&\text{if $\tfrac{\underline{\omega}}{\gamma}<\abs{\omega}<\tfrac{\overline{\omega}}{\gamma}$,}\\
1&\text{otherwise.}
\end{cases}
\end{equation}
Now let $\gamma_{k+1} = \tfrac{\overline{\omega}}{\underline{\omega}}\gamma_k$, where $\gamma_1=\overline{\omega}/\omega_{\mathrm{bw}}$, and set
\[
c_k=c_\varepsilon^{\gamma_k}.
\]
Observe that this construction ensures that $c_k\in\mathscr{S}_{\mathrm{bw}}$ (the case with $\gamma_1$ is easily observed from \cref{eq:sepfre}, and $\gamma_k\geq{}\gamma_1$). This construction also ensures that the frequency ranges with gains greater than 1 in \cref{eq:sepfre} can never overlap ($\overline{\omega}/\gamma_{k+1}=\underline{\omega}/\gamma_k$). Therefore for any $n\in\Nat$ and any $\omega\in\R$,
\[
\prod_{k=1}^n
\abs{\funof{\frac{pc_k}{1+pc_k}}\jw}\leq{}1+\varepsilon.
\]
This implies that for any $\varepsilon>0$,
\[
\inf_{c_1,c_2,\ldots{}\in\mathscr{S}}\sup_{n\in\Nat}\norm{\prod_{k=1}^n\frac{pc_k}{1+pc_k}}_\infty\leq{}1+\varepsilon{},
\]
which when combined with a) completes the proof.
\end{proof}

\begin{remark}\label{rem:1}
The assumption that all the $p_k$'s are identical with integrator dynamics is an assumption of convenience, and can be significantly relaxed. For example \Cref{thm:1} will continue to hold even if $p_k=s^{-m}q_k$, where $q_k$ is any minimum phase rational function (all the $q_k$'s may be different). It is conceivable that versions may continue to hold even if the $p_k$'s have right half-plane poles or zeros, though an infimal value of 1 may no longer be achievable, and the bandwidth limitation may need to be relaxed.
\end{remark}
\begin{remark}
Although the control laws used in the proof of \Cref{thm:1} meet a bandwidth constraint, the extreme parameter values used throughout render them unusable in practice (fortunately there are no infinitely long platoons of vehicles to control). However the basic approach used to construct the control laws, namely to mistune the frequency responses so as to mitigate the amplification of disturbances, may have some merit. A potentially profitable line of further investigation could be to investigate the scalability of control laws with simple parametrisations. For example if $m=2$, the PD control law
\[
c_k = 1 + k_ks
\]
could be considered, where $k_k$ is a random constant drawn from a known distribution (that could also possibly be designed). 
\end{remark}

\subsection{The bidirectional architecture}\label{sec:resb}

We now turn our attention to the bidirectional architecture from \cref{sec:prob}. The results in this subsection are considerably more restrictive than those presented for predecessor following, and are only really relevant when $p_k=s^{-1}$ or $p_k=s^{-2}$ (though these are arguably the two most significant values of $m$, c.f. \cite{TBS23}). They do however serve to illustrate the power of introducing only a single controller with dynamics different from the others when designing for scalable performance.

Our contribution is based on the following algebraic curiosity.

\begin{lemma}\label{lem:1}
Let $n\in\Nat$, $H_n=\diag{h_1,\ldots{},h_n}$, and
\[
  X_n =\scalemath{.8}{ \begin{bmatrix}
        1 & 0 & \cdots{} &\cdots&0 \\
        -1 & \ddots{} & \ddots{} &\ddots{} &\vdots{}\\
        0 & \ddots & \ddots & \ddots&\vdots{}  \\
        \vdots{} & \ddots{} & \ddots{}& \ddots & 0\\
        0 & \cdots{} & 0 & -1 &1
        \end{bmatrix} }\in \R^{n\times n}.
\]
If $h_n=\tfrac{1}{s}$ and $h_k=\tfrac{s+1}{s^2}$ for all $k<n$, then 
\[
I+X_nH_nX_n^{\top{}}=\tfrac{1}{s^2}U_nL_n
\]
where
\[
U_n=
\scalemath{.8}{\begin{bmatrix}
s+1&-1&0&\cdots{}&0\\
0&s+1&-1&\ddots{}&\vdots{}\\
\vdots{}&0&\ddots{}&\ddots{}&0\\
\vdots{}&\ddots{}&\ddots{}&\ddots{}&-1\\
0&\cdots{}&\cdots{}&0&s+1
\end{bmatrix}}
\]
and
\[
L_n=
\scalemath{.8}{\begin{bmatrix}
s&0&\cdots{}&\cdots{}&0\\
-1&s+1&0&\ddots{}&\vdots{}\\
0&-1&s+1&\ddots{}&\vdots{}\\
\vdots{}&\ddots{}&\ddots{}&\ddots{}&0\\
0&\cdots{}&0&-1&s+1
\end{bmatrix}}.
\]
\end{lemma}
\begin{proof}
It is readily seen that $\tfrac{1}{s^2}U_1L_1=1+X_1H_1X_1^{\top{}}$. Make the induction hypothesis that
\begin{equation}\label{eq:indhyp}
\tfrac{1}{s^2}U_nL_n=1+X_nH_nX_n^{\top{}}.
\end{equation}
Observe that $I+X_{n+1}H_{n+1}X_{n+1}^{\top{}}$ equals
\[
I+
\begin{bmatrix}X_n&0\\-e_n&1
\end{bmatrix}
\begin{bmatrix}
H_n+\bar{H}_n&0\\0&\tfrac{1}{s}
\end{bmatrix}
\begin{bmatrix}X_n^\top{}&-e_n^\top{}\\0&1
\end{bmatrix},
\]
where
\[
e_n = \begin{bmatrix}
0&\ldots{}&0&1
\end{bmatrix}\;\text{and}\;\bar{H}_n=\diag{0,\ldots{},0,\tfrac{1}{s^2}}.
\]
Therefore
\begin{equation}\label{eq:compit}
I+X_{n+1}H_{n+1}X_{n+1}^{\top{}}=
\begin{bmatrix}
I+X_{n}H_{n}X_{n}^{\top{}}&0\\0&0
\end{bmatrix}
+
M,
\end{equation}
where
\[
\begin{aligned}
M&=\begin{bmatrix}
X_{n}\bar{H}_{n}X_{n}^{\top{}}&-X_n\funof{H_n+\bar{H}_n}e_n^{\top{}}\\
-e_n\funof{H_n+\bar{H}_n}X_n^{\top{}}&1+e_n\funof{H_n+\bar{H}_n}e_n^{\top{}}+\tfrac{1}{s}
\end{bmatrix},\\
&=
\begin{bmatrix}
\tfrac{1}{s^2}e_n^{\top{}}e_n&-\tfrac{s+1}{s^2}e_n^{\top{}}\\-\tfrac{s+1}{s^2}e_n&\tfrac{\funof{s+1}^2}{s^2}
\end{bmatrix}.
\end{aligned}
\]
Making a similar calculation for $U_{n+1}L_{n+1}$ shows that
\[
\begin{aligned}
U_{n+1}L_{n+1}&=\begin{bmatrix}
U_{n}&-e_n^{\top{}}\\0&s+1
\end{bmatrix}\begin{bmatrix}
L_{n}&0\\-e_n&s+1
\end{bmatrix},\\
&=
\begin{bmatrix}
U_nL_n&0\\0&0
\end{bmatrix}
+
\begin{bmatrix}
e_n^{\top{}}e_n&-\funof{s+1}e_n^{\top{}}\\
-\funof{s+1}e_n&\funof{s+1}^2
\end{bmatrix}.
\end{aligned}
\]
Therefore by \cref{eq:indhyp}
\[
\tfrac{1}{s^2}U_{n+1}L_{n+1}=
\begin{bmatrix}
I+X_{n}H_{n}X_{n}^{\top{}}&0\\0&0
\end{bmatrix}
+
M.
\]
By comparison with \cref{eq:compit} we see that this implies that
\[
\tfrac{1}{s^2}U_{n+1}L_{n+1}=I+X_{n+1}H_{n+1}X_{n+1}^{\top{}},
\]
and so the result follows by induction.
\end{proof}

Our use of \Cref{lem:1} is centred on the following elementary observation about lower and upper triangular matrices. Given any lower triangular matrix $\bar{L}$ and any upper triangular matrix $\bar{U}$, whenever the following matrix product is invertible, it satisfies
\[
\funof{
\begin{bmatrix}
\bar{U}&\star\\0&\star
\end{bmatrix}
\begin{bmatrix}
\bar{L}&0\\\star{}&\star
\end{bmatrix}
}^{-1}=
\begin{bmatrix}
\bar{L}^{-1}\bar{U}^{-1}&\star{}\\\star{}&\star
\end{bmatrix}.
\]
The key point here is that the upper left block of the inverse of the product only depends on the upper left block of each of the factors. This means that if we consider the sequence of matrices
\begin{equation}\label{eq:matrixsens}
S_n=\funof{I+X_nH_nX_n^{\top{}}}^{-1},\;n=1,2,3,...
\end{equation}
the top-left entries in all these matrices will be the same. That is,
\begin{equation}\label{eq:inv1}
\funof{S_1}_{11}=\funof{S_2}_{11}=\funof{S_3}_{11}=\cdots{}.
\end{equation}
The same is true of the top-left $2\times{}2$ blocks of $S_2,S_3,S_4,\ldots{}$, the top-left $3\times{}3$ blocks of $S_3,S_4,S_5,\ldots{}$, and so on. 

Let's now interpret this observation in the context of the bidirectional model from \cref{sec:prob}. If we choose the control laws such that
\begin{equation}\label{eq:contl}
c_ks^{-m}=\begin{cases}
\tfrac{s+1}{s^2}&\text{if for $k=1,\ldots{}n-1$,}\\
\tfrac{1}{s}&\text{if $k=n$,}
\end{cases}
\end{equation}
then the matrices in \cref{eq:matrixsens} correspond precisely to the transfer functions from $d$ to $\Delta{}y$ in platoons with $n$ vehicles (see \Cref{fig:block}). The meaning of \cref{eq:inv1} is then that the closed loop transfer function from $d_1$ to $\Delta{}y_1$ is \textit{completely invariant} of the number of vehicles in the platoon. Since an analogous observation is true of the top-left blocks of any size, all of the closed loop transfer functions between elements of $d$ and $\Delta{}y$ are independent of the number of vehicles. And since the matrix sensitivity function $S_n$ appears in almost every closed loop transfer function in \Cref{fig:block}, every entry in almost every closed loop transfer function in the platoon becomes \textit{completely invariant} of $n$ too. And all that is needed to achieve this is to have the final vehicle in the platoon slightly adjust its control law relative to all the others according to \cref{eq:contl}.

\begin{figure}
\vspace{.2cm}
\input{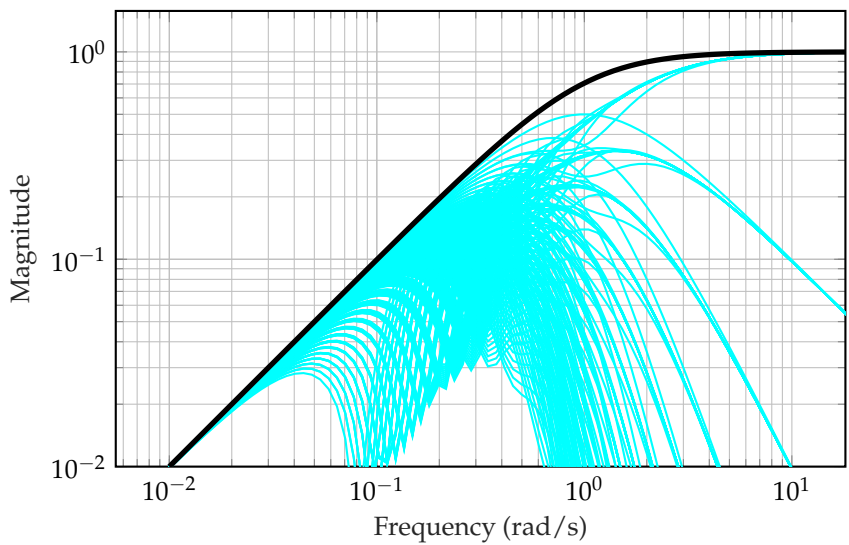}
\caption{The bode magnitude plot for every element of $S_{20}$. Every curve lies below the magnitude plot of $\tfrac{s}{s+1}$.}
\label{fig:bidpla}
\end{figure}

\begin{remark}
The entries of $S_n$ look like sensitivity functions designed using standard objectives (i.e. the control laws in \cref{eq:contl} are not unreasonable). This is easily seen in the top-left entry, which is given by
\[
\funof{S_n}_{11}=\frac{s}{s+1}.
\]
That is, $\abs{\funof{S_n}_{11}\jw}\ll1$ for low frequencies, with a maximum gain of 1. Different bandwidths can be obtained by making the substitution $s\mapsto{}s/T$, with $T>0$. Explicit expressions can be found for every entry in $S_n$, not just the top-left entry, by considering the expressions for the inverses of $U_n$ and $L_n$ (these inverses are very simple). Although they are higher order, they have the same qualitative shape as $\funof{S_n}_{11}$, as illustrated in \Cref{fig:bidpla}.
\end{remark}
\begin{remark}
Versions of \Cref{lem:1} with higher order dynamics suitable for analysing situations with higher values of $m$ are almost certainly possible. The presented arguments are more a proof of concept, than a completed story.
\end{remark}

\section{Conclusions}

This paper investigated the use of heterogeneous controller dynamics to enhance the scalability and performance of the ‘predecessor following' and ‘bidirectional' control architectures for platoons. In both cases heterogeneous controller dynamics made it possible to mitigate the issues of disturbance amplification and dependency on platoon length, allowing for improved performance in the large-scale setting. 

\bibliographystyle{IEEEtran}
\bibliography{references.bib}

\end{document}